\documentstyle[preprint,aps]{revtex}
\begin{document}
\draft
\preprint{\begin{tabular}{l}
\hbox to\hsize{June, 1997 \hfill SNUTP 97-026}\\[-3mm]
\hbox to\hsize{hep-ph/9706387 \hfill }\\[5mm] \end{tabular} }
\bigskip
\title{ Supersymmetric interpretation of high--$Q^2$ 
\\ HERA events 
and other related issues}
\author{Jihn E. Kim$^{a,b}$\thanks{jekim@huhepl.harvard.edu}~ 
and ~P. Ko$^c$\thanks{ pko@phyc.snu.ac.kr}\\ }
\address{$^a$ Lyman Laboratory of Physics, Harvard University\\
Cambridge, Massachusetts 02138\\
$^b$ Department of Physics,
Seoul National University\\
Seoul 151-742, Korea            \\
$^c$ Department of Physics, KAIST, Taejon 305-701, Korea \\
}
\date{\today}
\maketitle
 \tighten
\begin{abstract}
In the framework of the minimal supersymmetric standard model (MSSM) with 
$R-$parity violation, the high--$Q^2$ HERA events can be interpreted as 
the $s-$channel production of a single stop of $M_{\tilde{t}_1} \approx 200$ 
GeV, whose dominant   decay modes are assumed to be the $R-$parity violating 
$e^+ + d $ and the $R-$parity conserving $\chi^+ + b$. 
Assuming only one coupling $\overline{\lambda_{131}^{'}}$ is nonzero of
order $\sim 0.04-0.12$, we find that (i) the high$-Q^2$  HERA events 
can be understood as an $s-$channel stop production with a subsequent decay 
into $e^+ + ({\rm single ~jet})$, and   
(ii) the ALEPH 4-jet
events can be understood in the scenario suggested by
Carena {\it et al.}. We briefly discuss other physics signals of this 
scenario at other places such as HERA, LEP200 and Tevatron.
The best test for our scenario is to observe the stop decay into 
$\chi^+ + b$ followed by $\chi^+ \rightarrow \tilde{e}^+ + \nu_e $ and 
$\tilde{e}^+ \rightarrow 
q + \bar{q}^{'}$ via the $R-$parity violating coupling. 
\end{abstract}
\pacs{PACS : 13.60.-r, 11.30.Pb, 12.60.Jv, 13.10.+q}
\narrowtext





Recent observation of the high$-Q^2$ event excess in the $e^+ p$   
deep-inelastic scattering (DIS)  at HERA by both H1 and ZEUS Collaboration
may be a signal of new physics beyond the standard model (SM), 
if it is confirmed in the forthcoming run this year.
H1 reports 12 deep-inelastic events from data of 14 pb$^{-1}$ at 
$Q^2 > 15,000$ GeV$^2$ with 
large $e^+ -$jet invariant mass around $M \equiv \sqrt{xs} = 200 $ GeV 
\cite{h1}, where only  5  events are expected according to the standard
model (SM) prediction.
ZEUS reports 5 events from data of 20 pb$^{-1}$ at similar high $x$
and $Q^2 > 20,000~{\rm GeV}^2$  region where 2 events are expected
from the SM, although the events are scattered wider and thus the
resonance  structure in the Bjorken $x$ variable is not apparent \cite{zeus}.
These two data may not be consistent with each other\cite{drees} as well as 
with the SM prediction as a result of statistical fluctuations, and
we have to await   more data accumulation before we try to draw  any 
concrete conclusions  regarding the possible source of new
physics.  Combining the H1 and the  ZEUS data, one finds that 
\begin{equation}
\sigma ( e^+ + p \rightarrow e^+ + ({\rm a\ single\ jet}) + X ;\  Q^2 >
20,000  ~{\rm GeV}^2 ) \approx 0.2~{\rm pb},\nonumber
\end{equation}
whereas the SM contribution is negligible.  One simple way to increase
the neutral current (NC) cross section at HERA is to modify the parton
distribution functions (PDF's). The PDF's at very high $x$ and $ Q^2$ may be
altered so that one can have an excess of events of NC process at
HERA \cite{tung}. 
However, the possible resonance structure in $x \approx 0.44$
may be hard to explain in such attempts.  Also, PQCD corrections to
the NC process at HERA should be properly included in order to reduce
theoretical uncertainties within the SM \cite{rad}.
With these remarks kept in
mind,  it is tantalizing  to regard this HERA  anomaly  as a real
physics  signal, and to speculate what kind of new physics could
explain  this high--$Q^2$ HERA  events.  

There are many possibilities in choosing new physics scenarios beyond the SM.
Since the minimal supersymmetric standard model (MSSM) is one of the leading 
candidates for physics beyond the SM,
it provides us with a natural framework in which we can analyze any 
experimental anomalies that deviate  from SM predictions. 
The $R_b$ and $\alpha_{s} (M_{Z}^2)$ constituted such examples,
although these problems are gone now \cite{ichep96}.  
Another anomaly is the ALEPH four-jet events \cite{aleph} (which was not 
seen at other detectors, however).
Although the experimental situations need to be clarified among four groups 
at LEP, there have been some attempts to resolve the ALEPH 4-jet anomaly
in the framework of the MSSM with  $R-$parity violation \cite{godbole} 
\cite{carena}.  
In this work, we try to interpret the high--$Q^2$ HERA events in terms of
a single stop ($\tilde{t}_1$) production in the framework of the MSSM 
{\it with R-parity violation} \footnote{There is another model-indepedent 
approach based on the effective lagrangian language. In this framework, the 
high$-Q^2$ events at HERA can be fit to the following universal contact 
interaction lagrangian \cite{contact} :
 $ - {4 \pi \over \Lambda^2}~ \bar{e} \gamma^{\mu} \gamma_5 e~
\sum_{q=u,d,s} \bar{q} \gamma_{\mu} \gamma q $ for $\Lambda \simeq 3 $ TeV.
Such operator can be accomodated in theories with large family symmetry group,
$SU(12)$ \cite{nelson} or $SU(45)$ \cite{wyler}. 
However, such contact term seems to be against the other experiments from 
LEP2 \cite{opal} and the muonium hyperfine splitting \cite{ko}.}.  
More specifically,
we choose the scenario proposed by Carena {\it et al.} \cite{carena}
(the CGLW scenario) as a possible solution to the ALEPH 4-jet anomaly, 
because this scenario deals with pair production and decays of two particles 
with different masses at LEP  in the supersymmetric theories. 

In the MSSM, the $R-$parity violation is described by the following 
renormalizable superpotential,
\begin{equation}
W_{\not{R_{p}} } = \lambda_{ijk}~L_{i}L_{j} E_{k}^{c}
+ \overline{\lambda_{ijk}^{'}}~L_{i} Q_{j} D_{k}^{c}
+ \lambda_{ijk}^{''}~U_{i}^{c} D_{j}^{c} D_{k}^{c}
+ \mu_{i} L_{i} H_{2}.
\end{equation}
The proton decay can be avoided by setting $\lambda_{ijk}^{''} = 0$.  
In the CGLW scenario, it is assumed that only one of the couplings,  
$\overline{\lambda_{1jk}^{'}}$ 
( with $j,k = 1$ or 2 ),  is nonvanishing  
\footnote{This kind of assumption is certainly unnatural in  a sense.
However, in the presence of many $R-$parity violating couplings, it is worth
while to make such an assumption, and study its consequences at low energy
phenomena and at colliders, especially when we do not have any
theories that explain  the hierarchies in the fermion masses and the CKM
matrix elements.}
in the range of ${\rm (a~ few)} \times 10^{-4} < 
\overline{\lambda_{1jk}^{'}} < 10^{-2}$. 
Then, the ALEPH 4-jet events can be interpreted as
$ e^+ e^- \rightarrow \tilde{e}_L \tilde{e}_R $ via neutralino exchanges,
followed by the $R-$parity violating decay of the $\tilde{e}_L$ into two
jets. The other state, $\tilde{e}_R$, also decays into two jets
through the $\tilde{e}_L-\tilde{e}_R$ mixing \cite{carena}.
CGLW assumed that $M_{\tilde{e}_L} = 58$ GeV, $M_{\tilde{e}_R} = 48$ GeV,
in order to fit the dijet invariant masses reported by ALEPH Collaboration.
Also $M_{1} = 80-100$ GeV, and $M_{2} \ge 500$ GeV  in order to suppress
the sneutrino pair productions in the $e^+ e^-$ annihilations at LEP. 
With this choice of parameters, the lightest neutralino is dominantly bino 
$\tilde{B}$ \cite{carena}, with mass around 100--120 GeV.

The $\overline{\lambda^{'}}$ couplings in (2)
are unique in the sense that they  can be probed 
at HERA where $e^- (e^{+})$ and $u (d)$ in the proton can make
$\tilde{d}_{Rj}$ ($\tilde{u}_{Lj}$) resonance via the $R-$parity violating
$\overline{\lambda_{1j1}^{'}}$  coupling ($j=1,2$ or 3).
Therefore, the CGLW scenario can be tested at DESY if the spectrum of
superparticles satisfies certain conditions.  
There are various constraints on these couplings \cite{rp_low1}
\cite{rp_low2}.  It turns out that 
$\overline{\lambda_{131}^{'}}$ is less constrained than other couplings.  
Thus we assume that only $\overline{\lambda_{131}^{'}}$ is nonvanishing 
in (2). The most stringent limit on this coupling comes from the 
atomic parity violation (APV) \cite{rp_low1}. 
The most important contribution comes from the light stop exchange induced 
by (2) (or (7) below). From the new data on APV  
\cite{apv} \cite{pdg},  we have 
\begin{equation}
| \lambda_{131}^{'} \cos \phi_{\tilde{t}} | < 0.12
\end{equation}
for $M_{\tilde{t}_1} = 200$ GeV.  The $\tilde{t}_L-\tilde{t}_R$ mixing is 
taken into account in terms of the mixing angle $\phi_{\tilde{t}}$ :
\begin{eqnarray}
\tilde{t}_1 & = & \tilde{t}_L \cos \phi_{\tilde{t}}
 - \tilde{t}_R \sin \phi_{\tilde{t}},
\\
\tilde{t}_2 & = & \tilde{t}_L \sin \phi_{\tilde{t}} 
+ \tilde{t}_R \cos \phi_{\tilde{t}}.
\end{eqnarray}  
We have assumed $V_{33}^{\dagger} = 1$ so that $ \lambda_{131}^{'} \simeq 
\overline{\lambda_{131}^{'}}$. (See (6) below.)
This new limit (3) is almost a factor of four improvement compared to the 
older bound ($< 0.4$) obtained by Barger {\it et al.} \cite{rp_low1}.
Also, note that the above constraint on $\lambda_{131}^{'}$ is 
diluted by a factor of $\cos \phi_{\tilde{t}}$ as a result of 
the $\tilde{t}_L-\tilde{t}_R$ mixing. In our model, we assume that 
$\phi_{\tilde{t}} = \pi /4$, so that the above constraint  becomes
\begin{equation}
0.06 \leq | \overline{\lambda_{131}^{'}} |  \leq 0.17
\end{equation}
for $M_{\tilde{t}_1} = 200$ GeV.  The lower limit in (6) comes
from the requirement that this coupling is relevant to the HERA
high$-Q^2$ anomaly through the $s-$channel single stop production.
( See (18)--(20)  below.)

Assuming that the only nonvanishing coupling in (2) is 
$\overline{\lambda_{131}^{'}}$, let us write down the $R-$parity violating 
interaction lagrangian in terms of component fields :
\begin{eqnarray}
{\cal L}_{int, \not{R_{p}}} & = &
\overline{\lambda_{131}^{'}}~\left[ ~ \left(
\tilde{\nu}_{eL} \overline{d_{R}} b_{L}
+ \tilde{b}_{L} \overline{d_{R}}  \nu_{eL} + \tilde{d}_{R}^{*}
\overline{(\nu_{eL})^c} b_{L} \right)  \right.
\label{eq:int}
\\
& & \left. - V_{3j}^{\dagger} \left(~ \tilde{e}_{L} \overline{d_{R}} u_{jL}
+ \tilde{u}_{jL} \overline{d_{R}} e_{L} + \tilde{d}_{R}^{*}
 \overline{(e_{L})^c} u_{jL}~ \right) ~\right] + {\rm  h.c.}.   \nonumber
\end{eqnarray}
Note that a squark in the MSSM behaves like a scalar leptoquark if
$\lambda^{'} \neq 0$.  However, squarks  can also decay through the
$R-$parity conserving interactions, which distinguishes the squarks in
the MSSM with $R-$parity violation from the usual scalar LQ.  
All the fields in (7) are interaction eigenstates, and one has to 
consider the difference between these states and the mass eigenstates.
Since there is no theory of CKM mixing matrix yet, we do not know how
to relate the $U_{L,R}$ and $D_{L,R}$ with the $V_{CKM} \equiv V = 
U_{L}^{\dagger} D_L$ \cite{agashe}. 
In other words, only $V_{CKM}$ is known from
experiments, and we don't know $U_L$ and $D_L$ separately.  One can
have $U_L = 1$ or $D_L = 1$ in the extreme, but both $U_L $ and $D_L$
can differ from the unit matrix, either.  In this work, we assume that
$D_L = 1$ so that $V_{CKM} = U_L^{\dagger}$.
In (7), we have taken into account this flavor mixing effects in the
up--quark sector in terms of the CKM matrix elements $V_{jk}$ \footnote{Most 
other recent works put the flavor mixing in the down quark sector in terms of
the CKM matrix elements. In these cases, one cannot have $\lambda_{121}^{'}$ 
and $\lambda_{131}^{'}$ simultaneously, unlike our case. See the next 
paragraph.}, so that we have  the following induced interactions  :
\begin{equation}
\delta {\cal L}_{int, \not{R_{p}}} 
= - \lambda_{1j1}^{'} 
\left[~ \tilde{e}_{L} \overline{d_{R}} u_{jL}
+ \tilde{u}_{jL} \overline{d_{R}} e_{L} + \tilde{d}_{R}^{*}
 \overline{(e_{L})^c} u_{jL}~ \right] + {\rm  h.c.} ~~({\rm for}~ j = 1,2).
\end{equation}
For $\overline{\lambda_{131}^{'}} \cos \phi_{\tilde{t}} \sim 0.04-0.12$,
the coupling for the induced interaction vertex 
$\tilde{e}_{L} \overline{d_{R}} c_L$ is about 
$\lambda_{121}^{'} = V_{23} \overline{\lambda_{131}^{'}} \sim ( 2- 6 )
\times 10^{-3}$, which is in the  right order of magnitude suggested by CGLW
in order to solve the ALEPH anomaly.  
This is one of the key observations of this work, which has not 
been considered in other recent  works.  

The details of phenomenological aspects of our model presented in this work 
depend on the superparticle spectra, although the global features would be
generic. Since we are aiming at solving both the ALEPH four-jet 
anomaly and the high--$Q^2$ HERA events by a single $R-$parity violating
coupling $\overline{\lambda_{131}^{'}}$ in (7), we assume the same 
parameters  with the CGLW scenario : namely, 
\begin{itemize}

\item We assume the same parameters with the CGLW scenario, in order to 
solve the ALEPH 4--jet anomaly.
\begin{eqnarray}
&\tan \beta = 1,~~~~  
M_{\tilde{e}_L} = 58~{\rm GeV},~~~~M_{\tilde{e}_R} = 48~{\rm GeV},
\nonumber\\
&M_{1} = 80 ~{\rm GeV}, ~~~~ M_{2} = 500~{\rm GeV}. 
\end{eqnarray}
These conditions ensure that the $\tilde{e}_L \tilde{e}_R^{*} +
\tilde{e}_L^{*}  \tilde{e}_R $ production 
cross section at LEP2 is dominant over other channels, such as 
$e^+ e^- \rightarrow \tilde{\nu}_e \tilde{\nu}_e^{*}, \tilde{e}_L 
\tilde{e}_L^{*}, \tilde{e}_R \tilde{e}_R^{*}$.
Note that $M_1$ and $M_2$ are not related with each other through the
usual GUT relation \footnote{ The GUT relations on gaugino masses
were  imposed  in Ref.~\cite{carena2}, for example.}.  
Since $\tan \beta = 1$ in this model, the weak $SU(2)$ relation
\begin{equation}
M_{\tilde{\nu}_e}^2 = M_{\tilde{e}_L}^2 + ( 1 - \sin^2 \theta_{\rm w} ) \cos
(2\beta) M_{Z}^2, 
\end{equation}
implies that  $M_{\tilde{\nu}_e}^2 = M_{\tilde{e}_L}^2 = 58 $ GeV.
Also for $\tan \beta = 1$, the light neutral Higgs ($h^0$) mass entirely
arise from the radiative corrections, and $M_{h} < 80$ GeV. Therefore,
the light Higgs is within the reach of LEP2. 

\item Furthermore, we assume the universal squark masses
in order to do more concrete numerical analyses, and 
require the lighter stop mass is 200 GeV, assuming it is 
responsible for the high--$Q^2$ HERA events.
\begin{eqnarray}
&M_{\tilde{Q}} = M_{\tilde{U}} = M_{ \tilde{D}} = \tilde{m}, 
\nonumber\\
&M_{\tilde{t}_1} = 200~{\rm GeV},  
\end{eqnarray}
For the numerical value of $\tilde{m}$, we choose 250--300 GeV, so that 
the squarks (except for stops) have masses that could be probed at the 
Tevatron Upgrade.  Then heavier stop ($\tilde{t}_2$) has a mass around 440
GeV.  In this case, the SUSY contribution to $\Delta \rho$ is about 
$0.8 \times 10^{-3}$, and the lighter Higgs mass is about $62$ GeV.
\end{itemize}
For given $M_2 = 500$ GeV and $\tan\beta = 1$, 
the $\mu$ parameter is determined from the condition on the chargino mass
to be less than $M_{\tilde{t}_1} - M_b$ as well as the condition (18) below, 
which is obtained by saturating the high$-Q^2$ HERA events by the s-channel 
stop production and its  decay into $e^+ + d$.
For each $\mu$, we calculate the stop masses from the stop mass matrix in the
$( \tilde{t}_L, \tilde{t}_R )$ basis :
\begin{eqnarray}
M_{\tilde{t}}^2 & = & \left( \begin{array}{cc}
M_{\tilde{t}_L}^2 + M_{t}^2 + M_{Z}^2 ( {1\over 2} -
{ 2 \over 3} \sin^2 \theta_{\rm W} ) \cos 2 \beta  &  M_{t} ( A_{t} +
\mu \cot \beta )   \\
M_{t} ( A_{t} + \mu \cot \beta ) & M_{\tilde{t}_R}^2 + M_{t}^2 + {2 \over 3}
M_{Z}^2   \sin^2 \theta_{\rm W}  \cos 2 \beta
\end{array} \right)  
\\
& \longrightarrow & 
\left( \begin{array}{cc}
\tilde{m}^2 + M_{t}^2  
&  M_{t} ( A_{t} +
\mu )   \\
M_{t} ( A_{t} + \mu ) & \tilde{m}^2 + M_{t}^2 
\end{array} \right), 
\end{eqnarray}
so that $\phi_{\tilde{t}} = \pi/4$. 
We can get $A_t$ by requiring the lighter stop has the mass $M_{\tilde{t_1}} 
= 200 {\rm GeV} < \tilde{m}$, and $| A_{t} | < \tilde{m}$. 
The latter condition ensures that the ground state is color and charge neutral.
For each $\mu$, we can choose such $\tilde{m}$ which meets our above criteria.
Since this sector is not of our primary concern, we will not consider this
any further in this work. 



Now consider the $s-$channel stop (the lighter one, called 
$\tilde{t}_1$)  production cross section at HERA.  
Neglecting the SM contributions from $t-$channel exchanges of 
$\gamma$ and $Z$ bosons,  we have 
\begin{eqnarray}
&& {d^2 \sigma \over dx dy} ( e^+ (k,s) + p \rightarrow \tilde{t}_{1} 
\rightarrow e^+ (k^{'}, s^{'}) 
+ {\rm single~jet} ) \\
& = &
{4 \pi \over M_{\tilde{t}_1}^2}~{ s \Gamma^2 ( \tilde{t}_1 
\rightarrow e^+ d ) \over ( x s - M_{\tilde{t}_1}^2 )^2 + M_{\tilde{t}_1}^2
\Gamma_{\rm tot}^2 ( \tilde{t}_1 )  }~x d(x,Q^2)
\\
& \rightarrow &   {4 \pi^2 \over M_{\tilde{t}_1} \Gamma_{\rm tot} 
(\tilde{t}_1) }~
\Gamma^2 ( \tilde{t}_1 \rightarrow e^+ d ) ~d(x_{\rm res},Q^2)~\delta ( x s - 
M_{\tilde{t}_1}^2 ),
\end{eqnarray}
where $x = \hat{s} / s$,  $Q^2 = - ( k - k^{'})^2 = x y s $, 
$\Gamma_{\rm tot} ( \tilde{t}_1 )$ is the total decay rate for $\tilde{t}_1$,
and $\Gamma ( \tilde{t}_1 \rightarrow e^+ d )$  is the decay rate for   
the $R-$parity violating $\tilde{t}_1 \rightarrow e^+ d$ :
\begin{equation}
\Gamma ( \tilde{t_1} \rightarrow e^+ d ) 
= {m_{\tilde{t_1}} \over 16 \pi}~\cos^2 
\phi_{\tilde{t}}~| \lambda_{131}^{'} 
V_{33} |^2 \simeq 
39.8~{\rm MeV}~\left( { | \lambda_{131}^{'} ~\cos \phi_{\tilde{t}} | 
\over 0.10 } \right)^2.
\end{equation}
The last line Eq.~(16) is obtained in the limit of a narrow 
approximation for the stop resonance ($x_{\rm res} \equiv M_{\tilde{t}_1}^2 
/ s_{\rm HERA}$).
In this limit, the cross section $\sigma ( e^+ + p \rightarrow \tilde{t}_1 
\rightarrow e^+ + d + X )$ becomes proportional to 
the  branching ratio $B(\tilde{t}_1 \rightarrow 
e^+ d )$ times the decay rate for the same mode. 
In other words,  the  cross section for $ e^+ + p \rightarrow \tilde{t}_1 
\rightarrow e^+ + d + X $ is proportional to 
$B ( \tilde{t}_1 \rightarrow e^+ d ) \times | \lambda_{131}^{'} \cos 
\phi_{\tilde{t}} |^2$.  
Using the CTEQ3 PDF \cite{cteq3} and assuming $\sigma (e^+ p
\rightarrow \tilde{t}_1 \rightarrow e^+ + $ (single jet)) = 0.2 pb as a
representative value of the high$-Q^2$ HERA events, we get  
\begin{equation}
\sqrt{B} \times | \lambda_{131}^{'} \cos \phi_{\tilde{t}} | \approx 0.04,
\end{equation}
which agrees with the others' results.   Then, the condition $B \leq
1$  implies that
\begin{eqnarray}
0.04 \leq  | \lambda_{131}^{'} \cos \phi_{\tilde{t}} | \leq 0.12
\\
B (\tilde{t}_1 \rightarrow e^+ d ) \geq 0.11.
\end{eqnarray}

In the presence of nonvanishing $\lambda_{131}^{'}$ coupling, the stop behaves
like a leptoquark that  can decay into $e^+ + d$. 
Therefore the scalar leptoquark (LQ) search at HERA and at the Tevatron via 
$LQ \rightarrow e + $ (single jet) decay mode gives some
constraints on the properties of the stop. 
The D0 limit on the stop mass is 225 GeV for $B(LQ \rightarrow e + 
{\rm jet}) = 100 \%$ \cite{d0}, when 
the QCD corrections to the LQ pair prodution at the Tevatron is included 
\cite{zerwas}.  CDF limit is $> 210$ GeV for the same case \cite{cdf},  
and a stop of $M_{\tilde{t}_1} = 200$ GeV can be safely accomodated 
if $B ( \tilde{t}_1 \rightarrow  e^+ + d) < 70 \%$. 
In other words, we need an extra decay channel of the stop other than the 
$R-$parity violating decay, $\tilde{t}_1 \rightarrow e^+ d$.  
At this point, it is worthwhile to remember  that  there are 
differences between the stop and the conventional scalar leptoquark.
First, the stop does not couple to  $\bar{\nu} + u$. Secondly, the stop 
can have $R-$parity conserving decay modes in addition to the $R-$parity 
violating mode. For example, one can imagine a stop decay 
into a chargino ($\chi^+$) plus a bottom quark, 
if the chargino is light enough.  
Therefore, we shall assume that the lighter chargino state $\chi_1$, one
of the eigenstate of 
\begin{eqnarray}
M_{\tilde{\chi^{\pm}}}  =  \left(   \begin{array}{cc}
M_{2} & \sqrt{2} M_{W} \sin \beta  \\
\sqrt{2} M_{W} \cos \beta & - \mu
\end{array}     \right)
 \rightarrow  \left(   \begin{array}{cc}
M_{2} &  M_{W} \\
M_{W} & - \mu
\end{array}     \right), 
\end{eqnarray}
has mass below $M_{\tilde{t}_1} - M_{b}$ so that
the decay $\tilde{t}_1 \rightarrow \chi_1^+ + b$ is 
kinematically allowed. On the other hand, the decay into 
$t + \chi_1^0$ is kinematically forbidden.  
This condition constrains a possible range of the $\mu$ parameter for a given
$M_2$ and $\tan \beta$.  

The  decay rate of this mode has been calculated by Kon {\it et al.} 
\cite{kon} :
\begin{eqnarray}
\Gamma ( \tilde{t}_1 \rightarrow b \chi^+_{k} ) & = & 
{\alpha \over   4 \sin^2 
\theta_{\rm W} M_{\tilde{t}_1}^3}~\lambda^{1/2} ( M_{\tilde{t}_1}^2, M_{b}^2,
M_{\chi_k}^2)    \nonumber 
\\
& \times & \left[ \left( \left| G_{L} \right|^2 + \left| G_{R} \right|^2 
\right) ( M_{\tilde{t}_1}^2
-  M_{b}^2 - M_{\chi_k}^2 ) - 4 M_{b} M_{\chi_k}~ {\rm Re}~
\left( G_{R} G_{L}^{*} \right) \right],  
\end{eqnarray}
where 
\begin{eqnarray}
G_{L} & = & - {M_{b} U_{k2}^{*} \cos \phi_{\tilde{t}}
 \over \sqrt{2} M_{W} \cos \beta},
\\
G_{R} & = & V_{k1} \cos \phi_{\tilde{t}} 
+ {M_{t} V_{k2} \sin \phi_{\tilde{t}} \over \sqrt{2} 
M_{W} \sin \beta}.  
\end{eqnarray}
In the above equations, $V_{ij}, U_{ij}$ are matrices that diagonalize 
the chargino mass matrix (21), viz. 
$ U^{*} M_{\tilde{\chi}^{\pm}} V^{-1} = M_{D}$.  

If charginos are heavier than the stop, we have to consider $R-$parity 
conserving decay via virtual chargino exchange : 
\begin{eqnarray*}
\tilde{t}_1 \rightarrow b + \chi^{+*} 
& \rightarrow & b + (e^+ + \tilde{\nu}_e) , 
\\
& & b + (\tilde{e}^+ + \nu_e).
\end{eqnarray*}
If the lightest neutralino ($\chi_{1}^0$) is light enough,  we have to 
consider $\tilde{t}_1 \rightarrow b W^+ \chi_{1}^0$ as well.  
This is indeed the case in our framework, since we assume $M_{\tilde{t}_1} 
= 200$ GeV, and  $M_{\chi_1^0} \approx 100-120$ GeV.  
We have also analyzed these three--body decay modes in the limit of a 
pure bino neutralino ($\chi_1^0 \approx \tilde{B}$) \cite{hikasa}, 
and found that they are all
negligible (smaller than 0.1 MeV) compared to the two--body decay modes
discussed above : $R-$parity violating decay  $\tilde{t}_1 \rightarrow
e^+  + d$ and $R-$parity conserving decay $\tilde{t}_1 \rightarrow b +
\chi_{1}^+$.  Therefore we consider only these two two-body decays of 
$\tilde{t}_1$ in the following.  

In Figs.~1 (a) and (b), we show the branching ratio for $\tilde{t}_1 
\rightarrow e^+ + d$ as a function of $\mu$ for $\mu > 0$ and $\mu < 0$,
respectively. We chose five different values of $\lambda_{131}^{'} \cos 
\theta_{\tilde{t}} = 0.04$ (the solid curve), 0.06, 0.08, 0.10 and 0.12 
(the long dashed curve). 
In Fig.~2, we show the $\mu$ dependence of the cross section for 
$e^+ + p \rightarrow \chi_1^+ + b + X$  for both $\mu > 0$ (the solid curve) 
and $\mu < 0$ (the dashed curve) for the maximally allowed $\lambda_{131}^{'} 
\cos\theta_{\tilde{t}} = 0.12$, assuming $M_2 = 500$ GeV and $\tan \beta 
= 1$ \footnote{Because of constraints (17) and (19), the $\mu$ parameter 
is fixed upto a sign for a fixed $\lambda_{131}^{'}$.  
However, the cross section for 
$e^+ + p \rightarrow e^+ + $(single jet) may be  changed in the future 
after more data are accumulated. Therefore we do not
impose these conditions in Figs.~1 (a),(b) and Fig.~2.}.
We have included the $t-$channel $\tilde{\nu}$ exchange diagram \cite{kon} 
(induced by the $R-$parity violating coupling in  (7))  
as well as the $s-$channel stop production  diagram. 
We note that the cross section  is smaller than $0.1$ pb for most 
regions of $\mu$ parameters, so that it is unlikely that this mode is 
a useful   place to test our scenario at the $e^+ p$ collider.
If we chose different $(M_2, \tan \beta)$, then the corresponding 
production cross section would change accordingly.  
  
In Tables~1 and 2, we choose six different values of $\lambda_{131}^{'} \cos 
\phi_{\tilde{t}}$, and find  $\mu$ parameters which satisfies the conditions 
(17) and (19) for $\mu > 0$ and $\mu < 0$, respectively.
The resulting  chargino masses $(M_{\chi_1}, M_{\chi_2})$,
the decay rates for $\tilde{t}_1 \rightarrow  \chi_1^+ + b$ and $\tilde{t}_1 
\rightarrow e^+ + d$, and the total decay width of the stop are given in 
these tables, neglecting the three--body decay modes of the stop.  
In the last row of Tables~1 and 2, we also list the cross section for 
$e^+ + p \rightarrow \chi_1^+ + b$ 
for $x> 0.1$ and $Q^2 > 15,000$ GeV$^2$. Here, $Q^2 \equiv -(k - k^{'})^2$,
and $k$ and $k^{'}$ are the 4-momenta of the initial positron and 
the final chargino, respectively.
We note that the cross section for $e^+ + p \rightarrow \chi_1^+ + b + X$
is rather small ($< 0.1$ pb), which indicates that this channel may not be
detected soon.  

Finally let us consider the decay rates of the lightest chargino $\chi_1^+$.
Its main decay modes are 
\begin{itemize}
\item Two--body decays : 
$\chi_1^+ \rightarrow \tilde{e}^+ + \nu_e,~e^+ + \tilde{\nu}_e$ 
\item Three--body decays :
$\chi_1^+ \rightarrow \chi_{1}^0 + \bar{l} \nu_e, ~\chi_{1}^0 + q \bar{q}^{'}$.
\end{itemize}
It turns out that the dominant decay mode for  $M_2 = 500$ MeV (and $\mu$ 
considered in Figs.~1 and 2) is 
$\chi_1^+ \rightarrow \tilde{e}^+ + \nu_e$, with its width being around 
several hundred MeV which is quite broad.  
Other decay modes are all negligible compared with  this  channel.
The final SUSY particles eventually decay into ordinary particles through 
$R-$parity violating couplings : 
\begin{itemize}
\item $\tilde{e} \rightarrow q + \bar{q}^{'}$ 
\item $\tilde{\nu}_e \rightarrow \bar{b} + d$
\item $\chi_1^0 \rightarrow \tilde{e}^{+} +  e, \tilde{e} + e^+ , 
\tilde{\nu}_e + \nu_e.$
\end{itemize}
Therefore, the signal for the $e^+ + p  \rightarrow 
\chi_1^+ + b$ would be multijets with missing energy accompanying a soft 
$b-$jet. However, for our choice of $M_2$ and $\mu$, the cross section for 
this channel (with $x>0.1$ and $Q^2 > 15,000~{\rm GeV}^2$) 
is too small, $< 0.1$ pb, which makes almost hopeless to observe
it at HERA in the near future.  



As a further check to this scenario, let us note that 
the $t-$channel stop exchange can modify the 
$e^+ e^- \rightarrow d \bar{d}$ by interference with the SM constributions 
from the $s-$channel $\gamma, Z$ exchanges. 
This  effect may be observed at LEP2 upto $\sqrt{s} = 190$ GeV.  
In Fig.~3, we show the deviation of the cross section from the SM
value for $e^+ e^- \rightarrow d \bar{d}$
as a function of $\sqrt{s}$ upto $\sqrt{s} = 200$ GeV with 
 $\lambda_{131}^{'} \cos \phi_{\tilde{t}} = 0.12 $ 
(the solid curve) and 0.04 (the dashed curve), respectively.  
Deviations from the SM prediction is less than $- 1$ \% (destructive 
interference with the SM contribution \footnote{We thank J. Kalinowski for 
pointing  out this fact to us.}) at LEP2 for the maximal 
$\lambda_{131}^{'} \cos \phi_{\tilde{t}} = 0.12$, 
which is hardly discernable unless the integrated luminosity at LEP2 becomes 
larger than $O(1) fb^{-1}$.

The $t-$channel stop exchange also contributes to 
the Drell--Yan (DY) prodution of the $e^+ e^-$ (but not $\mu^+ \mu^-$)
pair at the Tevatron through the parton level subprocess, 
$d \bar{d} \rightarrow e^+ e^-$.  
Using the CTEQ3 PDF \cite{cteq3} and including the valence quark 
contributions only, we obtained  the $e^+ e^-$ invariant mass spectrum
at the Tevatron (with $\sqrt{s} = 1.8 $ TeV and the rapidity cut $|y|<1$),
along with the SM prediction in Fig.~4.  
We have included the heavier stop ($\tilde{t}_2$) effect as well.
The stop exchange enhances the DY yield of the $e^+ e^-$ pair at large
invariant mass $M_{ee}$. However, the 
difference between the SM prediction (the dashed curve) and our
model  stop (the solid curve) with the maximal value of
the coupling $\lambda_{131}^{'} \cos\phi_{\tilde{t}} = 0.12 $ 
is  at the level of a few \% or less because of the small coupling
$\lambda_{131}^{'}$, so that it would be impossible to test our model
via the DY production at the Tevatron.  However the leptoquark search
at the Tevatron will be cover some part of $B( \tilde{t}_1 \rightarrow
e^+ + d )$ for $M_{\tilde{t}_1} = 200$ GeV.  



In summary, we assumed that $R-$parity violation in the MSSM occurs through 
only one $R-$parity violating coupling $\overline{\lambda_{131}^{'}}$ in 
the interaction basis,   and that it induces effective couplings
$\lambda_{121}^{'}$ and $\lambda_{111}^{'}$ in the mass eigenstates 
by the flavor mixing among the up quark sector (see (8)).
If $\overline{\lambda_{131}^{'}} ~\cos\phi_{\tilde{t}} \sim 0.04-0.12$,
the induced $\lambda_{121}^{'}$ and $\lambda_{111}^{'}$ lie in the 
range where the ALEPH 4-jet events find a solution in terms of $e^+ e^-
\rightarrow \tilde{e}_L \tilde{e}_R$ and subsequent decays of the selectrons 
into $ u + \bar{d}$ (and its charge conjugate state) {\it \'{a} la} CGLW.   

We also considered implications of such scenario at other colliders
such as LEP and the Tevatron.  Deviations from the SM predictions for
the cross section  for $e^+ e^- \rightarrow  d \bar{d}$ and the DY yield 
of $e^+ e^-$ at the Tevatron are at most a
few \% or less, and it would not be easy to detect the virtual stop
effects at such colliders.  One possibility is to discover such stop
of $M_{\tilde{t}} = 200$ GeV at the Tevatron for large $B( \tilde{t}_1
\rightarrow e^+ d)$.  It would be very  interesting to see
if the high$-Q^2$ HERA events survive more data accumulation this year  
and if the production cross section at HERA for $e^+ + p \rightarrow
 \chi_1^+ + b$ ( followed by $\chi_1^+ \rightarrow
\tilde{e}^+ + \nu_e $ and  $\tilde{e}^+ \rightarrow 
q + \bar{q}^{'}$ via the $R-$parity violating coupling) is in the same
order of magnitude given in this work.  

A few remarks are in order before closing,  regarding the ALEPH 4--jet
anomaly and the high$-Q^2$  HERA events which were considered
simultaneously  in the present work.  We have assumed that the ALEPH
4--jet  anomaly {\it was real} and could be solved in terms of the
$R-$parity  violating interaction {\it \'{a} la} CGLW.
In case that the ALEPH 4--jet events disappear in the  future, overall
features of our scenario would not change very much, except that we
can have  different values of the gaugino masses, $M_1$ and $M_2$,
from those adopted in Ref.~\cite{carena}.  We also assumed that the
universal squark masses in this work, so that the $\tilde{t}_L -
\tilde{t}_R$  mixing angle was $\phi_{\tilde{t}} = \pi / 4$. 
This mixing angle can change if the universal squark masses do not hold.
Still, qualitative features of our predictions will not change,
unless $\phi_{\tilde{t}} = \pi /2$ (an extreme).

{\it Note added}
While we were preparing this work, there appeared several papers 
\cite{hera_mssm} which tried to interprete the high$-Q^2$ HERA events in
terms of  the MSSM with $R-$parity violation.  
However, there  are  some differences  in the choice of the SUSY
parameters,  the sparticle spectra and the decay channels of the stop.
Leptoquark scenarios were considered in Refs.~\cite{lepto}, and other 
approaches can be found in Refs.~\cite{others}.

\acknowledgements

The authors are grateful to Dr. M. Drees, Dr. S.B. Kim and Prof. 
S.K. Park for useful discussions. 
They also thank Dr. Jungil Lee for helping use the REDUCE. 
This work was supported in part by KOSEF through CTP at Seoul National
University, by the Hoam Foundation (JEK), by the Distinguished Scholar 
Exchange Program of Korea Research Foundation, by KOSEF (PK), 
and by the Ministry of Education Project No. BSRI--96--2418.


%

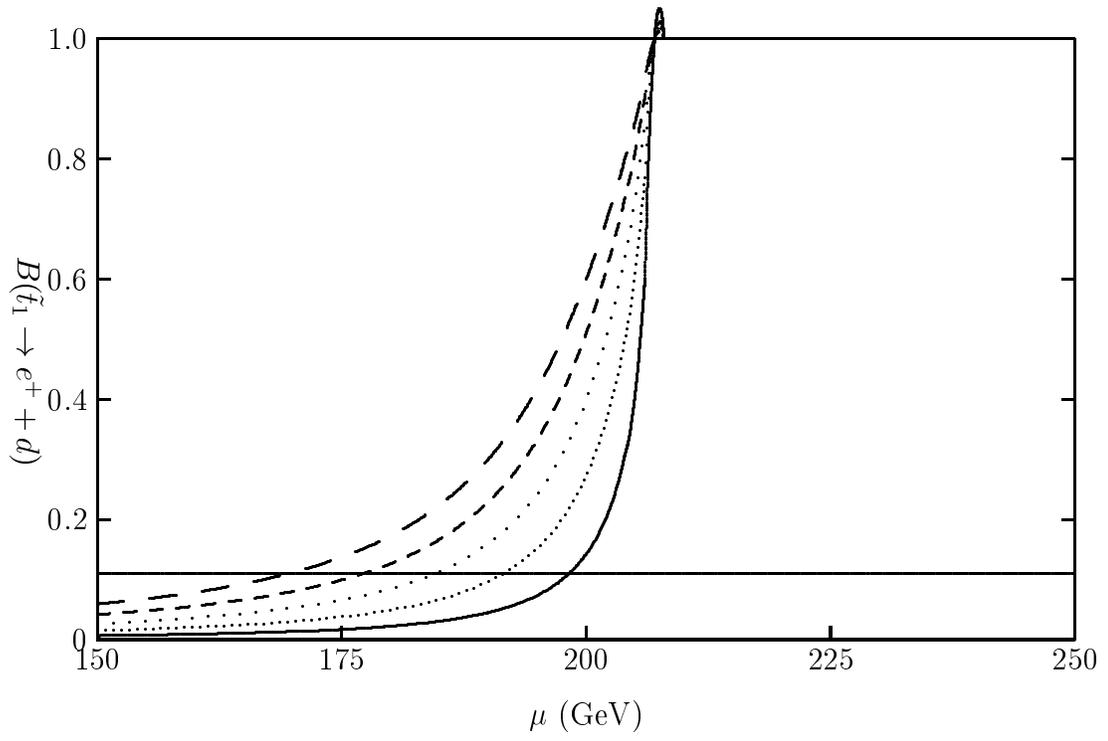
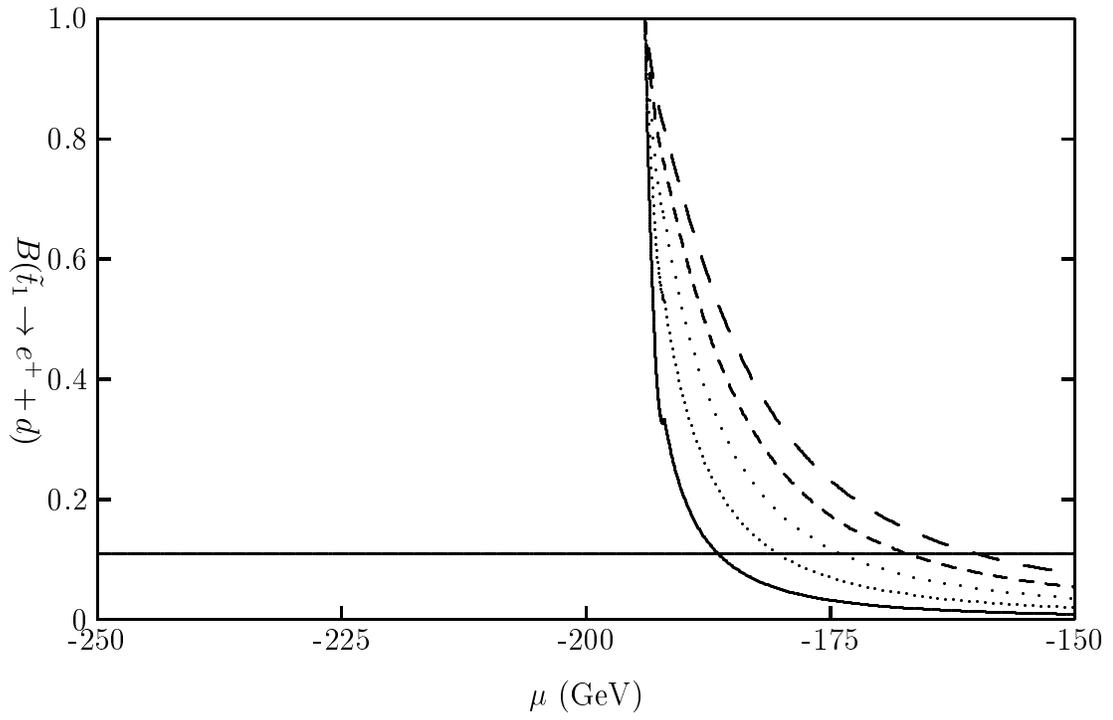
\begin{figure}
\caption{The branching ratio for $\tilde{t}_1 \rightarrow e^+ + d$
as a function of the $\mu$ parameter for $M_2 = 500$ GeV : 
(a) for $\mu>0$, and (b) for $\mu<0$. The five curves corresponds to 
$\lambda_{131}^{'} \cos \phi_{\tilde{t}} = 0.12$ (long dashed),
$0.10,0.08,0.06$, and  $\lambda_{131}^{'} \cos \phi_{\tilde{t}} = 
0.04$ (solid), respectively.}
\label{fig1}
\end{figure}

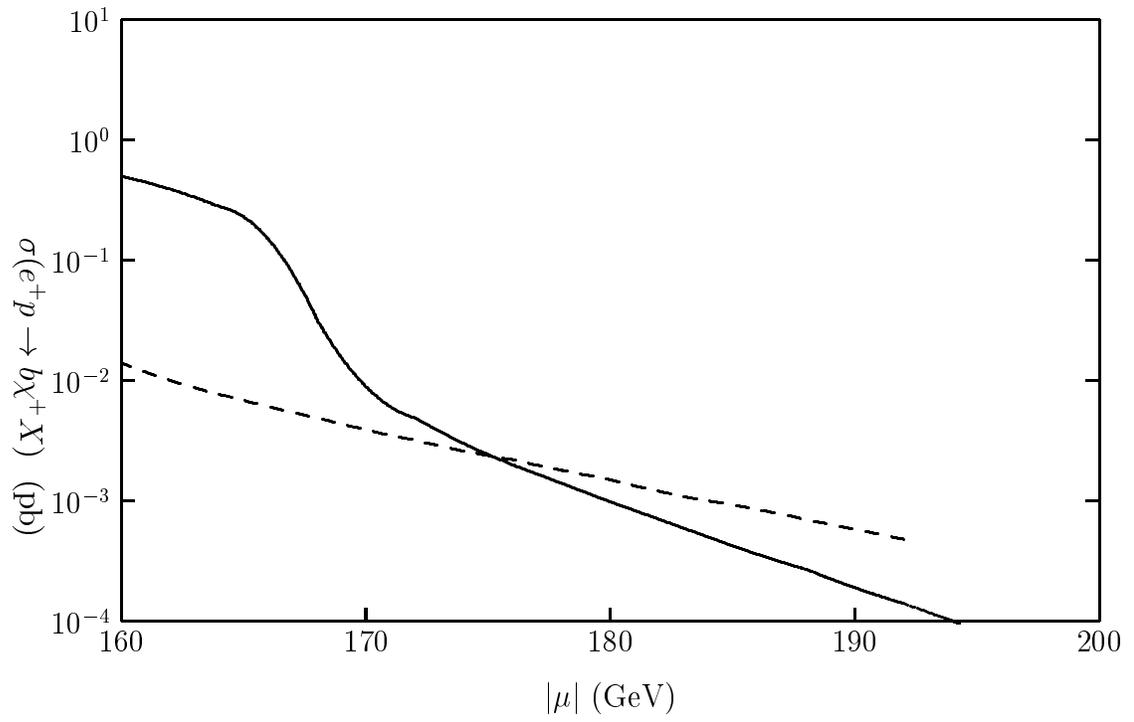
\begin{figure}
\caption{Production cross section for $e^+ + p \rightarrow  b + \chi_1^+ + X$
(including the $t-$channel $\tilde{\nu}_e$ exchange with $M_{\tilde{\nu}_e} = 
58$ GeV)  as a function of the $\mu$ parameter for $x > 0.1$ and 
$Q^2 > 15,000$ GeV$^2$.
We choose $M_2 = 500$ GeV, and the maximal allowed value of 
$\lambda_{131}^{'} \cos \phi_{\tilde{t}} = 0.12$.  The solid and the dashed 
curves correspond to $\mu > 0$ and $\mu < 0$, respectively.  }
\label{fig2}
\end{figure}

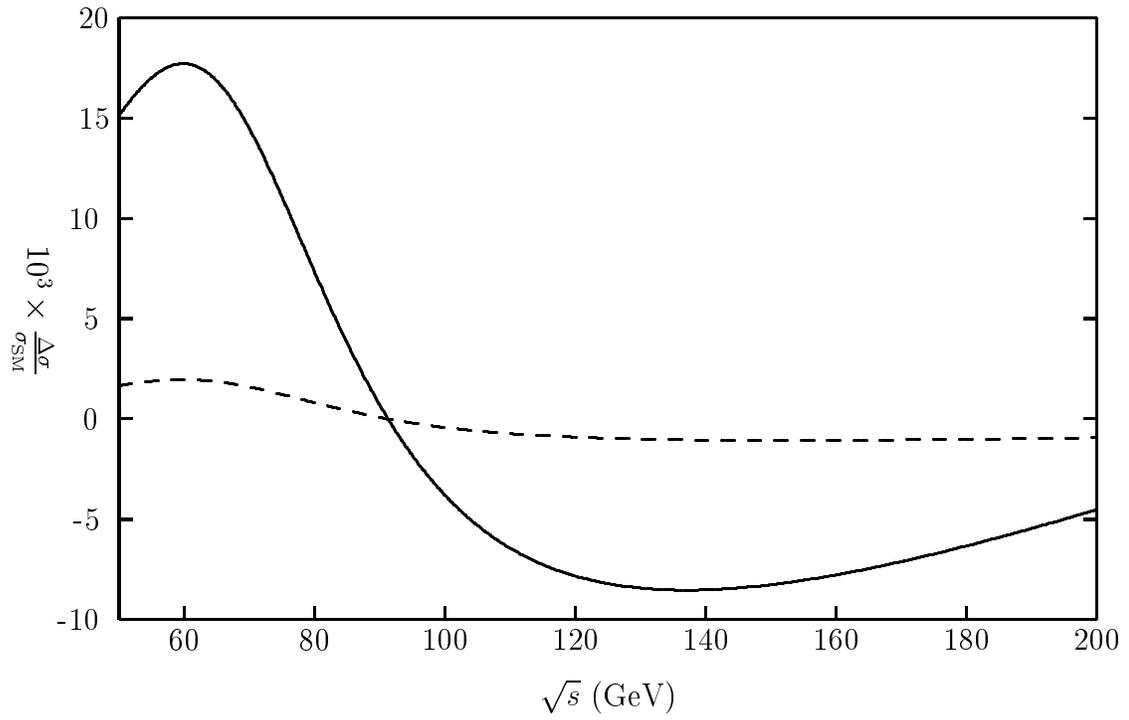
\begin{figure}
\caption{
Fractional deviations of the cross section for $e^+ e^- \rightarrow 
d \bar{d}$ from the SM prediction ($\Delta \sigma \equiv 
\sigma_{\rm (SM+SUSY)} - \sigma_{\rm (SM)}$ ) as a function of $\sqrt{s}$  : 
the stop contribution ($M_{\tilde{t}_1} = 200$ GeV)  with 
$\lambda_{131}^{'} \cos \phi_{\tilde{t}} = 0.12$ (the solid) and 0.04 
(the dashed), respectively. 
} 
\label{fig3}
\end{figure}

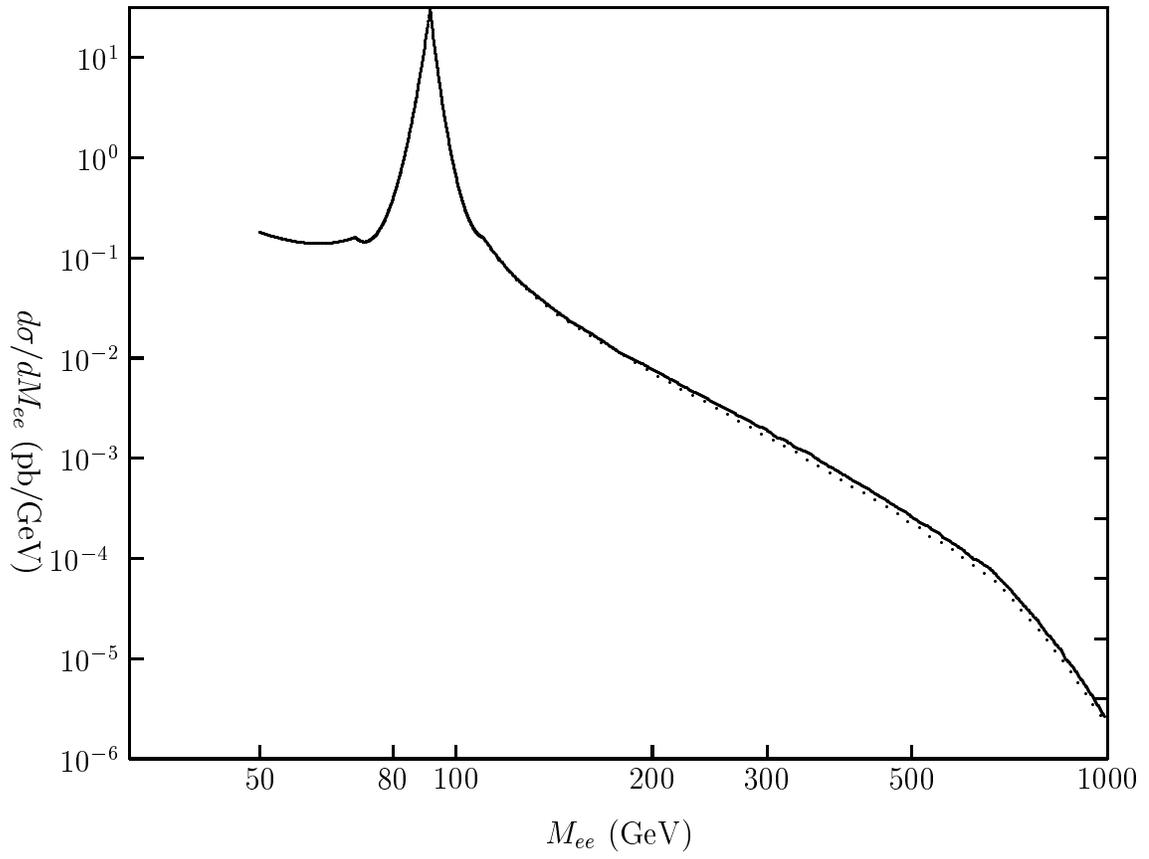
\begin{figure}
\caption{The Drell--Yan prodution of the $e^+ e^-$ pair at the Tevatron for 
the SM (the dashed) and the stop exchange with $\lambda_{131}^{'} \cos
\phi_{\tilde{t}} = 0.12$ (the solid).  We imposed the rapidity cut
$|y| < 1$, but ignored the radiative QCD corrections.}
 \label{fig4}
\end{figure}
 
%
%
\begin{table}
\caption{The $\mu (>0)$ parameter, the chargino masses, the decay rates for
$\tilde{t}_1$ into $e^+ + d$ and $\chi_1^+ + b$, and the cross section for 
$\chi_1^+ + b$ production at HERA including the $t-$channel $\tilde{\nu}_e$ 
(with $x > 0.1$ and $Q^2 > 15,000~{\rm GeV}^2$)  
for six different values of $\lambda_{131}^{'}$. We set $M_{\tilde{\nu}_e} = 
58$ GeV, $M_{2} = 500$ GeV, $\tan \beta = 1$, $\phi_{\tilde{t}} = \pi/ 4$,
and $M_{b} = 5.3$ GeV.}
\label{table1}
\begin{tabular}{c|ccccc}
$\lambda_{131}^{'} \cos\phi_{\tilde{t}}$ & 0.04 &  0.06 & 0.08 & 
0.10 & 0.12
\\ \hline
$\mu$ (GeV) & 207 & 203 & 195 & 184  &  170
\\  
$M_{\chi_1} $  (GeV)  &  195.2 & 191.4 & 183.8 & 173.3  & 159.9
\\
$ M_{\chi_2} $ (GeV)  & 517.2 & 517.1 & 516.8 & 516.3  & 515.9   
\\
$\Gamma ( \tilde{t}_1 \rightarrow e^+ d )$ (MeV) 
 & 6.4  & 14.3 & 25.5 & 39.8 & 57.3
\\
$\Gamma ( \tilde{t}_1 \rightarrow  \chi_{1}^+ + b) $ (MeV) 
& 0.0 & 19.5  & 80.0 & 212.7 & 453.8
\\
$\Gamma_{\rm tot}(  \tilde{t}_1 )$ (MeV) 
& 6.4  & 33.8 & 105.1 & 252.5 & 511.1
\\
$\sigma ( \chi_1^+ + b )$ (pb) & 0.0 & $0.67 \times  10^{-5}$ & $0.40 \times 
10^{-4}$ & $ 0.35 \times 10^{-3}$ & $ 0.89 \times 10^{-2}$
\end{tabular}
\end{table}

\vspace{.5in}

\begin{table}
\caption{The $\mu (<0)$ parameter, the chargino masses, the decay rates for
$\tilde{t}_1$ into $e^+ + d$ and $\chi_1^+ + b$, and the cross section for 
$\chi_1^+ + b$ production at HERA including the $t-$channel $\tilde{\nu}_e$ 
exchange (with $x > 0.1$ and $Q^2 > 15,000~
{\rm GeV}^2$) for six different values of $\lambda_{131}^{'}$. We set 
$M_{\tilde{\nu}_e} = 58$ GeV, $M_{2} = 500$ GeV, $\tan \beta = 1$, 
$\phi_{\tilde{t}} = \pi/ 4$, and  $M_{b} = 5.3$ GeV.}
\label{table2}
\begin{tabular}{c|ccccc}
$\lambda_{131}^{'} \cos\phi_{\tilde{t}}$ & 0.04 &  0.06 & 0.08 &
0.10 & 0.12
\\ \hline
$\mu$ (GeV) &  $ -194$ & $-191$ & $-184$  & $-174$  &  $-160$
\\
$ M_{\chi_1} $ (GeV)   & 195.0 & 191.8  & 185.0 & 175.3 & 161.7 
\\
$M_{\chi_2} $ (GeV)  & 512.0 & 512.3 & 512.3 & 512.2 & 512.1   
\\
$\Gamma ( \tilde{t}_1 \rightarrow e^+ d )$ (MeV)
& 6.4  & 14.3 & 25.5 & 39.8 & 57.3
\\
$\Gamma ( \tilde{t}_1 \rightarrow  \chi_{1}^+ + b) $ (MeV)
& $ 0 $ & $18.5 $  & 75.6   & 205.7 & 467.1
\\
$\Gamma_{\rm tot}(  \tilde{t}_1 )$ (MeV) &  6.4  & 32.8  & 101.1 &
245.5 & 524.4
\\
$\sigma ( \chi_1^+ + b )$ (pb) & 0.0 & $0.13 \times 10^{-3}$ & $0.45 \times 
10^{-3}$ & $0.18 \times 10^{-2}$ & $0.14 \times 10^{-1}$
\end{tabular}
\end{table}

\end{document}